\documentclass[prd]{revtex4}
\usepackage{graphicx}
\usepackage{dcolumn}
\usepackage{bm}
\def\alt{\mathrel{\hbox{\rlap{\hbox{\lower4pt\hbox{$\sim$}}}\hbox{$>$}}}}
\begin{document}

\title{Gravitational collapse in 
anti de Sitter space}

\author{David Garfinkle}
\email{david@physics.uoguelph.ca}
\affiliation{Department of Physics, University of Guelph, Guelph, Ontario,
Canada N1G 2W1
\\ and Perimeter Institute for Theoretical Physics, 
35 King Street North, Waterloo Ontario, Canada N2J 2W9}

\begin{abstract}
A numerical and analytic treatment is presented here of 
the evolution of initial data 
of the kind that was conjectured 
by Hertog, Horowitz and Maeda to lead to a violation 
of cosmic censorship.
That initial data is essentially a thick domain wall connecting two regions
of anti de Sitter space.  
The evolution results in no violation of cosmic censorship, 
but rather the 
formation of a small black hole.  

\end{abstract}
\maketitle

\section{Introduction}

A longstanding issue in general relativity is that of cosmic censorship:
the question of whether the singularities formed in gravitational 
collapse are hidden inside black holes.  This old question was given a
new twist by Hertog, Horowitz and Maeda\cite{Gary} who conjectured 
that cosmic censorship can be violated in certain spacetimes that
are asymptotically anti de Sitter.  The proposed counterexample 
consists of a matter model, an initial data set and an argument that
the evolution of that initial data gives rise to a singularity that 
cannot be hidden inside a black hole.  

The matter model is a scalar field with a potential with two minima
below zero.  The potential is chosen so that the system just barely
satisfies the positive mass theorem.  The initial data is essentially
a spherically symmetric, thick domain wall that interpolates between
regions of the two anti de Sitter spaces corresponding to the two
potential minima.  The data is chosen so that for an inner region of
radius $R$, the mass is proportional to $R$.  Since anti de Sitter
space is unstable, it is argued in\cite{Gary} that the evolution of
this initial data will result in a singularity in the central region
and further that for sufficiently large $R$ the mass of the spacetime
will not allow the formation of a black hole large enough to cover
the singularity.     

The nature of this issue was changed when Dafermos\cite{dafermos}
proved that any singularities formed in the evolution of the 
system of \cite{Gary} cannot be visible to observers at infinity.
Thus if the arguments of \cite{Gary} were correct, then the 
singularity would extend to infinity.  In other words the entire
space would collapse in a Big Crunch.

However, various authors\cite{me,Steve,Miguel,Miguel2} have
expressed misgivings about the arguments of \cite{Gary} 
(including the authors of \cite{Gary} in \cite{Gary2}).
The authors of \cite{me,Steve,Gary2} express doubts over the
assumption in \cite{Gary} that the central region can be well
approximated as a spacetime that is homogeneous but not anti de Sitter.
The authors of \cite{Miguel,Miguel2} raise the possibility that 
the wall might move outward indefinitely. 

It is not clear how to settle this issue using analytical means.
Therefore it makes sense to perform numerical simulations of the 
evolution of the initial data of \cite{Gary} and find the outcome.
Such a simulation was reported in\cite{me} for
$R=7$ with the result that a small black hole formed rather than a 
naked singularity.  However, this did not resolve the issue since it
was argued in\cite{Gary,Gary2} that ``sufficiently large'' $R$ 
means $R \ge 600$.  

In this work we simulate the evolution of
the initial data of\cite{Gary} for large
$R$.  Here too, the result is the formation of a small black hole,
not a naked singularity.  Though the results are numerical, the most
important properties of this system can be understood analytically
using the properties of perturbations of anti de Sitter space.

Sec. II presents the relevant equations and numerical method.  Sec.
III contains a treatment of perturbations of anti de Sitter space.
Results of the numerical simulations are presented in Sec. IV and 
conclusions in Sec. V. 

\section{Equations and numerical methods}

The system to be studied is a spherically symmetric scalar field $\phi$
with a potential $V$.  The appropriate equations are therefore the
Einstein-scalar equations:
\begin{eqnarray}
{G_{ab}} &=&  {\nabla _a} \phi {\nabla _b} \phi - {g_{ab}}
( {\textstyle {1 \over 2}} {\nabla ^c}\phi {\nabla _c} \phi + V )  
\label{einstein}
\\
{\nabla _a}{\nabla ^a} \phi &=& {{\partial V}\over {\partial \phi}}
\label{wave}
\end{eqnarray}
(Here we are using units where $8 \pi G =1$).
We use polar-radial coordinates for the metric which puts it in the form
\begin{equation}
d {s^2} = - {\alpha ^2} d {t^2} + {a^2} d {r^2} + {r^2} \left ( d 
{\theta ^2} + {\sin ^2} \theta d {\varphi ^2} \right )
\label{metric}
\end{equation}
It is helpful to define the quantities $X$ and $Y$ given by
$ X \equiv \partial \phi/\partial r$ and $Y \equiv (a/\alpha )
\partial \phi /\partial t$.   Then equation (\ref{wave}) yields the
following evolution equation for $Y$
\begin{equation}
{{\partial Y} \over {\partial t}} = {1 \over {r^2}} {\partial \over 
{\partial r}} \left ( {r^2} {\alpha \over a} X \right ) - \alpha a 
{{\partial V} \over {\partial \phi}}
\label{dty}
\end{equation}
while Einstein's equation yields the following equations which are used
to find the metric components $\alpha$ and $a$.
\begin{eqnarray}
{{\partial a}\over {\partial r}} &=& {{a(1-{a^2})}\over {2 r}} + 
{\textstyle {1 \over 4}} r a \left ( {X^2} + {Y^2} + 2 {a^2} V \right )
\label{dra}
\\
{\partial \over {\partial r}} \ln (a\alpha ) &=& {r \over 2} \left ( 
{X^2} + {Y^2}\right ) 
\label{draalpha}
\end{eqnarray}
Einstein's equation also provides an evolution equation for $a$.  Defining
the quantity ${\cal C} \equiv \partial a/\partial t - r \alpha X Y/2$
then it follows from Einstein's equation that $\cal C$ vanishes.  We
will use this as a code check.  That is, equations (\ref{dty}-\ref{draalpha})
are used to simulate the evolution of the system and then the results of that
simulation are checked by seeing that $\cal C$ converges to zero.

The numerical method used is essentially that of reference\cite{me}.  
That is spatial derivatives are replaced by centered differences for
unequally spaced points, while time evolution is done using the iterated
Crank-Nicholson (ICN) method.\cite{ICN}  The equations are stabilized
using Kreiss-Oliger dissipation.\cite{KO}  

The spacing of the grid points is different from that of\cite{me}.  Define
the coordinate $\rho$ by $r\equiv \tan \rho$.  Though the coordinate
$r$ is used in all equations, we choose a grid of points
that is equally spaced in $\rho$.  
 
The potential used is that of reference\cite{Gary}
\begin{equation}
V(\phi) = - 3 + 50 {\phi ^2} - 81 {\phi ^3} + k {\phi ^6}
\end{equation}
where the constant $k$ is chosen so that the system just barely satisfies 
the positive mass theorem.  The initial data is chosen to minimize the
contribution of the potential to the total mass.  
Here the minimum is found over all field configurations that are
in the true vacuum at $r=0$ and the false vacuum for $r \ge R$.
This leads to an ordinary
differential equation for $\phi$ that is solved using a shooting 
method as described in\cite{me}.  For our purposes, an important 
property of the solution is that in the central region 
$\phi \propto {r^\beta} $ where 
$\beta =({\sqrt {409}}-3)/2 \approx 8.6$.

\section{anti de Sitter perturbations}

We now consider the behavior of the central region as a perturbation
of anti de Sitter space.  The initial data in the central region
has $\phi \propto {r^\beta}$ where $\beta \approx 8.6$.  Thus the
scalar field is very small.  Keeping only terms up to linear order
in $\phi$ we find that equations (\ref{dra}-\ref{draalpha}) yield 
${\alpha ^2}={a^{-2}}=1+{r^2}$, while equation (\ref{dty}) becomes
\begin{equation}
- {{{\partial ^2}\phi}\over{\partial {t^2}}} + {{1+{r^2}}\over {r^2}}
{\partial \over {\partial r}} \left ( {r^2} [1+{r^2}] {{\partial \phi}
\over {\partial r}}\right ) - 100 (1+{r^2}) \phi = 0
\end{equation}
Defining $\psi \equiv r \phi$ and $\rho \equiv {{\tan}^{-1}} r $
we obtain
\begin{equation}
- {{{\partial ^2}\psi}\over {\partial {t^2}}} + 
{{{\partial ^2}\psi}\over {\partial {\rho ^2}}} - {{102 \psi}\over 
{{\cos ^2}
\rho}} = 0
\label{waverho}
\end{equation}
Provided that the spatial and temporal variation of $\psi$ is sufficiently 
large, the last term in this equation can be neglected yielding for
an ingoing wave $\psi = f(t+\rho )$ or
\begin{equation}
\phi = {1 \over r} f(t+ {\tan ^{-1}} r )
\label{packet}
\end{equation}

We will later see from the numerical results that for large walls,
the last term in equation (\ref{waverho}) is negligible.  However, 
the reason for this is easy to understand analytically:  Define a    
``coordinate thickness'' $\Delta r$ of the wall as follows:
Let $\Delta r$ be the amount that $r$ varies as $\phi$ varies
from $10\%$ to $90\%$ of its maximum value.  Correspondingly define
$\Delta \rho$.  Since the initial data depends only on $r/R$, it 
follows that $\Delta r \propto R$.  However, this means that for
large walls ({\it i.e.} those with $ R\gg 1$) we have
$\Delta \rho \propto {R^{-1}}$.   
Thus viewed as a function of $\rho$, the initial data is essentially
zero in the central region and then steeply increases to its
maximum value in a narrow region near $\pi /2$.  The evolution
of such initial data is a narrow wave packet that propagates inward.

Once the wave packet reaches values of $r \ll 1$, the difference
between anti de Sitter space and Minkowski space becomes irrelevant.
The subsequent behavior of the wave packet then depends on whether it
will shrink to a size smaller than its Schwarzschild radius or whether
it will disperse.

\section{numerical results}

All runs were done in double precision on a Sun Blade 2000.  Let $N+1$ be 
the number of spatial grid points.  We first present results for $R=100$.
Figure 1 shows the initial data for $\phi$.  The evolution of this initial
data produces very narrow wave packets; so all plots of the evolution of
the initial data will be confined to the range of $r$ where the wave packet
is present.  
\begin{figure}
\includegraphics[scale=0.8]{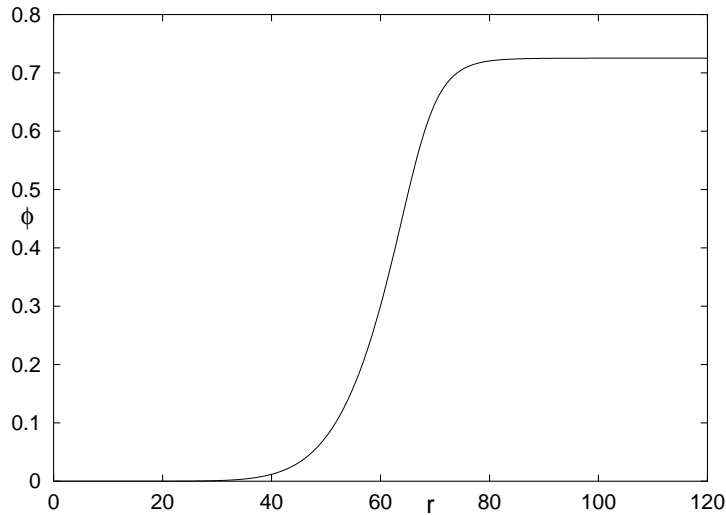}
\caption{\label{fig1} the scalar field $\phi$ at $t=0$.  $R=100$ and
$N=256000$}
\end{figure}

\begin{figure}
\includegraphics[scale=0.8]{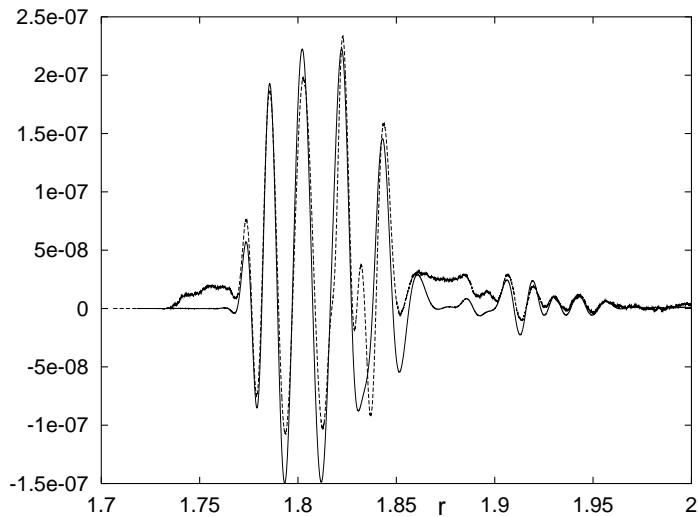}
\caption{\label{fig2} the constraint at $t=0.5$ for $R=100$ 
at two different resolutions:
$\cal C$ for $N=256000$ (solid line) and $8{\cal C}$ for $N=512000$
(dashed line)} 
\end{figure}

\begin{figure}
\includegraphics[scale=0.8]{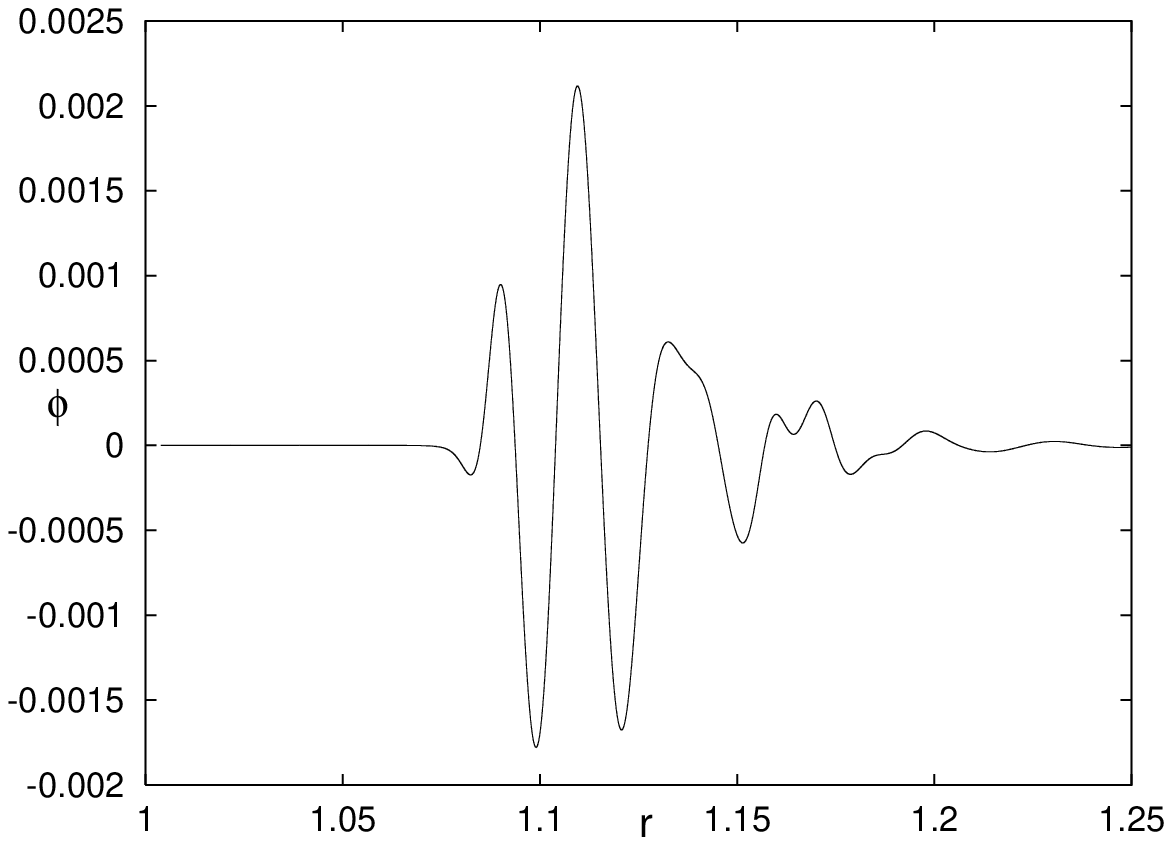}
\caption{\label{fig3} the scalar field $\phi$ at $t=0.731$.  $R=100$ and 
$N=256000$}
\end{figure}

\begin{figure}
\includegraphics[scale=0.8]{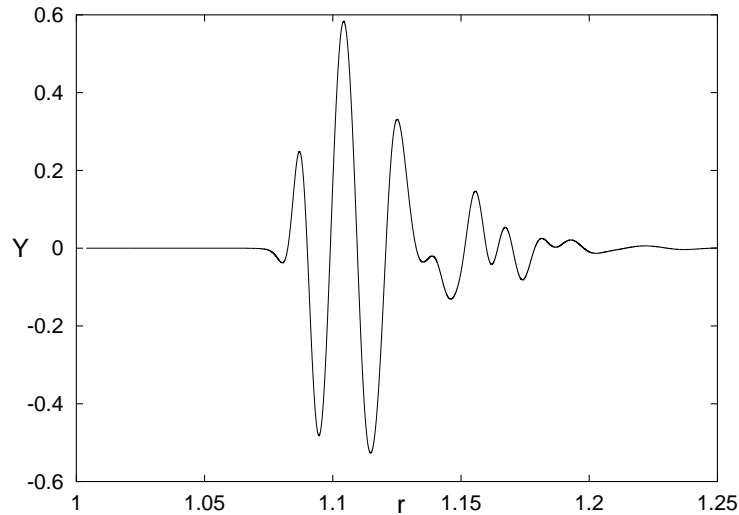}
\caption{\label{fig4} $Y$ at $t=0.731$. $R=100$ and $N=256000$}
\end{figure}
Figure 2 shows the result of a convergence test of the code.  Here, the
system is evolved to a time of 0.5 in two different runs: one with
$N=256000$ and the other with $N=512000$.  What is plotted in the figure
is the quantity $\cal C$ for the $N=256000$ run (solid line) and 
$8{\cal C}$ for the $N=512000$ run (dashed line).  
The result shows that the constraint converges to zero at third order.
This is a bit surprising, since from the finite difference techniques
used one would expect the code to be only second order convergent.  
However, it may be that for this type of ingoing wavepacket the 
leading part of the truncation error for the constraint vanishes.

Figures 3-9 show the results of a run with $N=256000$.
Figures 3 and 4 show respectively $\phi$ and $Y$ for
the evolution of the initial data to a time t=0.731.   
Note that $\partial \phi /\partial t \sim
570 \phi$.  Therefore, the last term in equation (\ref{waverho}) is
negligible, justifying the expression in equation (\ref{packet}).  
Figure 5 shows $r\phi$ as a function of $t+\rho$ for two times:
$t=0.2$ (solid line) and $t=0.4$ (dashed line).  These curves agree,
showing that equation (\ref{packet}) is a good approximation for this
part of the evolution.  In contrast, figure 6 shows $\phi$ as a function
of $r$ for these two times.  Here one sees two well separated wave
packets.

\begin{figure}
\includegraphics[scale=0.8]{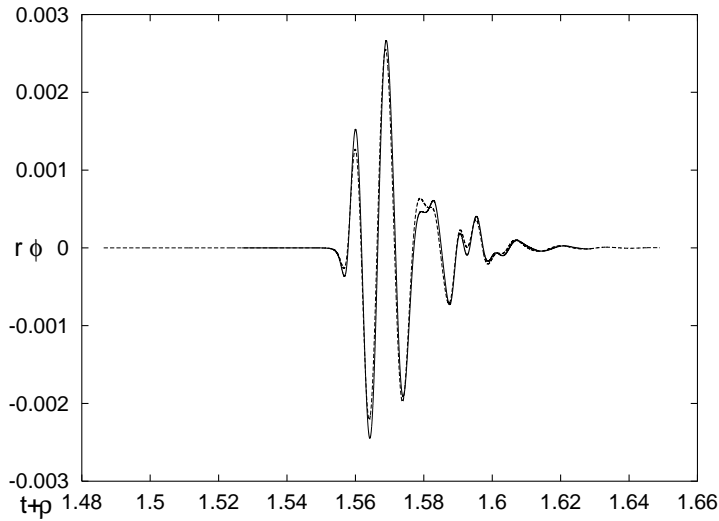}
\caption{\label{fig5} $r\phi$ as a function of $t+\rho$ at
$t=0.2$ (solid line) and $t=0.4$ (dashed line). $R=100$ and $N=256000$}
\end{figure}

\begin{figure}
\includegraphics[scale=0.8]{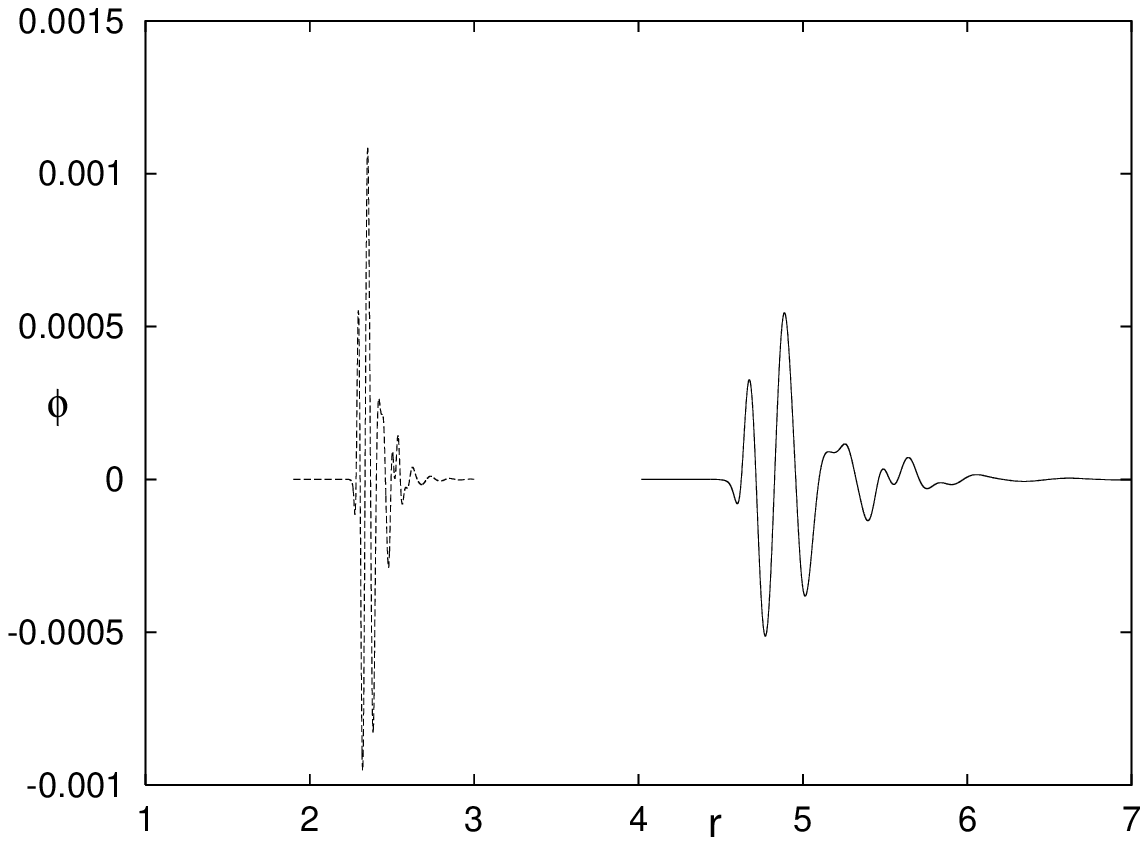}
\caption{\label{fig6} $\phi$ as a function of $r$ at
$t=0.2$ (solid line) and $t=0.4$ (dashed line). $R=100$ and $N=256000$}
\end{figure}

\begin{figure}
\includegraphics[scale=0.8]{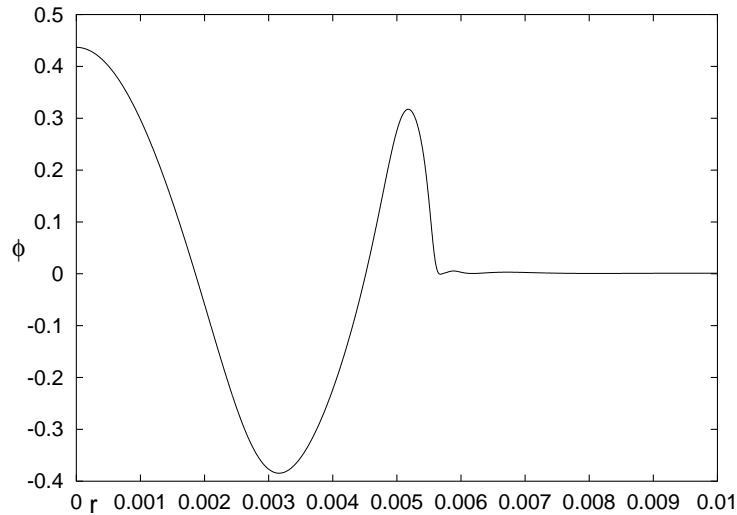}
\caption{\label{fig7} $\phi$ at the final time.
$R=100$ and $N=256000$}
\end{figure}

\begin{figure}
\includegraphics[scale=0.8]{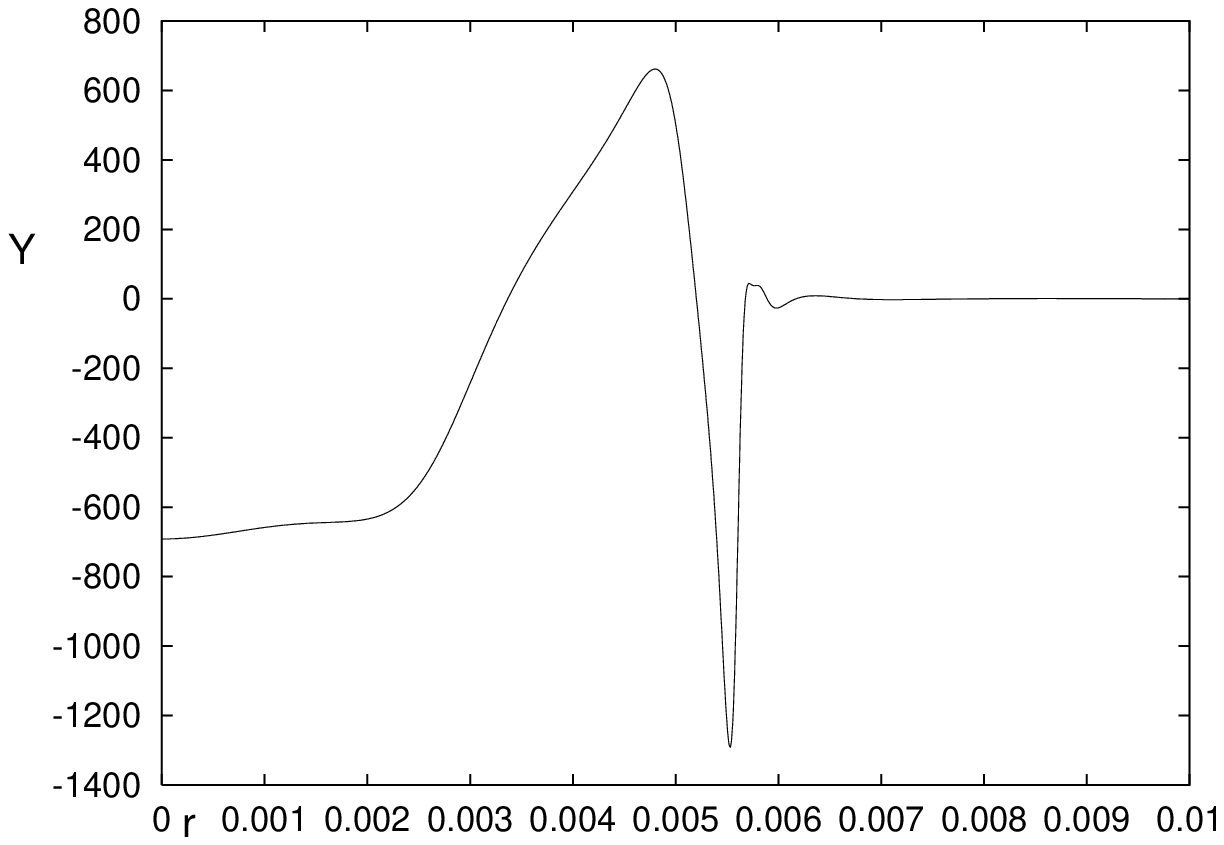}
\caption{\label{fig8} $Y$ at the final time.
$R=100$ and $N=256000$}
\end{figure}

\begin{figure}
\includegraphics[scale=0.8]{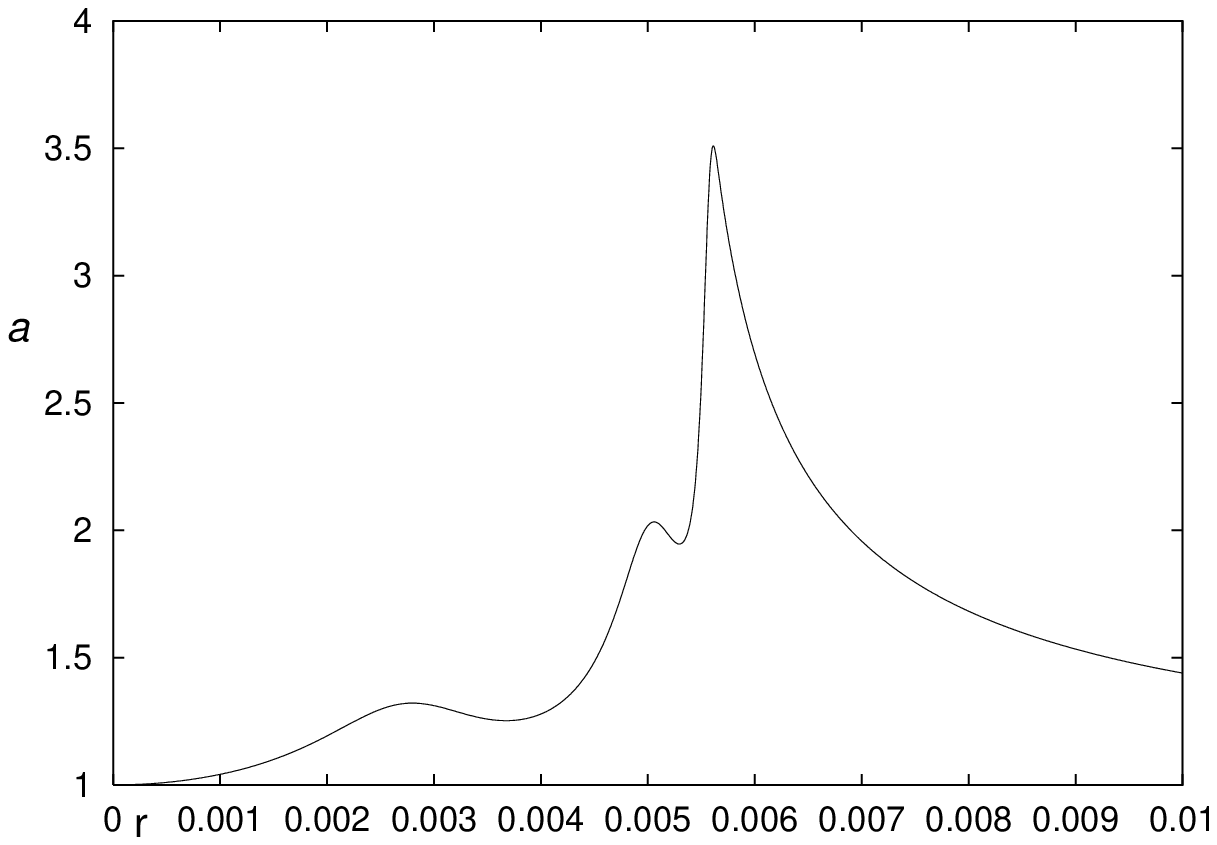}
\caption{\label{fig9} $a$ at the final time.
$R=100$ and $N=256000$}
\end{figure}

Figures 7, 8 and 9 show respectively $\phi , \; Y$ and $a$ at the 
final time.  Here the final time, which is slightly less than 
$\pi /2$, is chosen by having the simulation end when the maximum
value of $a$ reaches 3.5 corresponding to 
$2m/r=0.92$.  The reason for this is that the 
coordinate system used here breaks down when a trapped surface forms
and that breakdown is signalled by $a$ becoming large where the trapped 
surface forms.  Thus figure 9 indicates that a trapped surface forms
at $r \approx 0.0056$, while figures 7 and 8 indicate that the region of
high curvature is contained within this trapped surface.  Thus the
result of the evolution of the initial data with $R=100$ is a small
black hole.  

\begin{figure}
\includegraphics[scale=0.8]{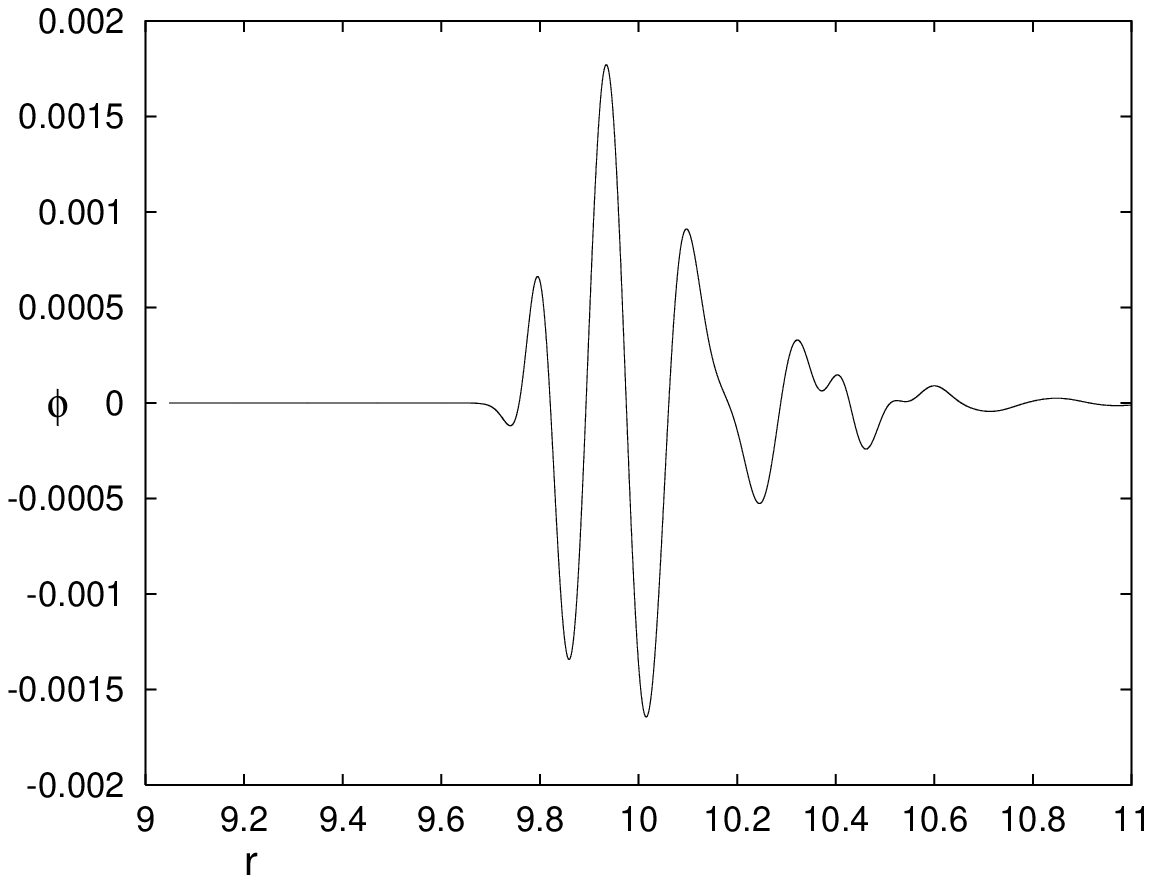}
\caption{\label{fig10} $\phi$ at $t=0.1$. $R=600$ and $N=1024000$}
\end{figure}

\begin{figure}
\includegraphics[scale=0.8]{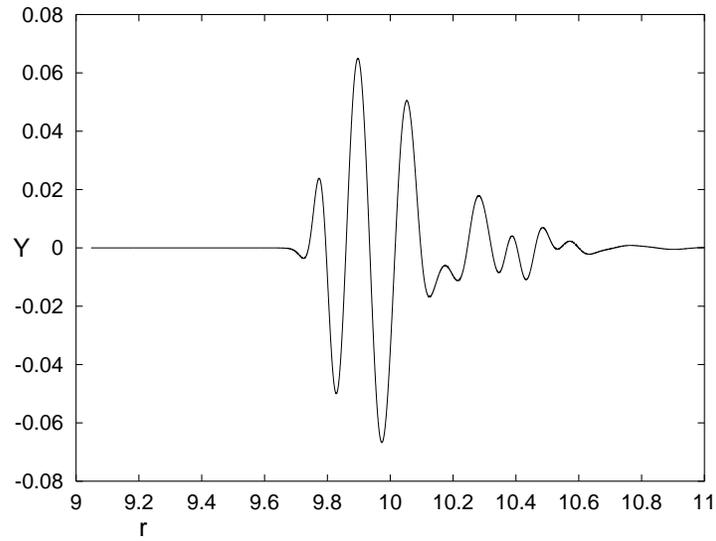}
\caption{\label{fig11} $Y$ at $t=0.1$. $R=600$ and $N=1024000$}
\end{figure}

\begin{figure}
\includegraphics[scale=0.8]{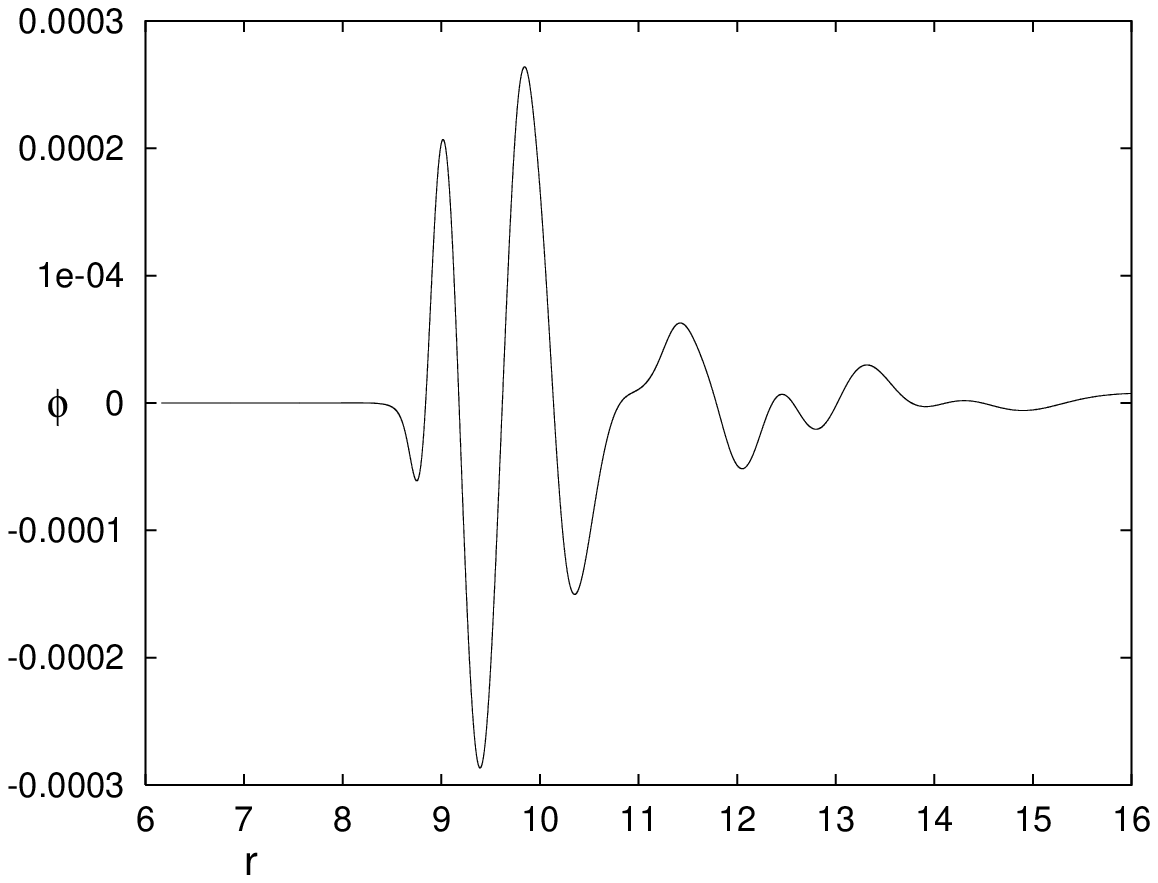}
\caption{\label{fig12} $\phi$ at $t=0.1$. $R=100$ and $N=256000$}
\end{figure}

\begin{figure}
\includegraphics[scale=0.8]{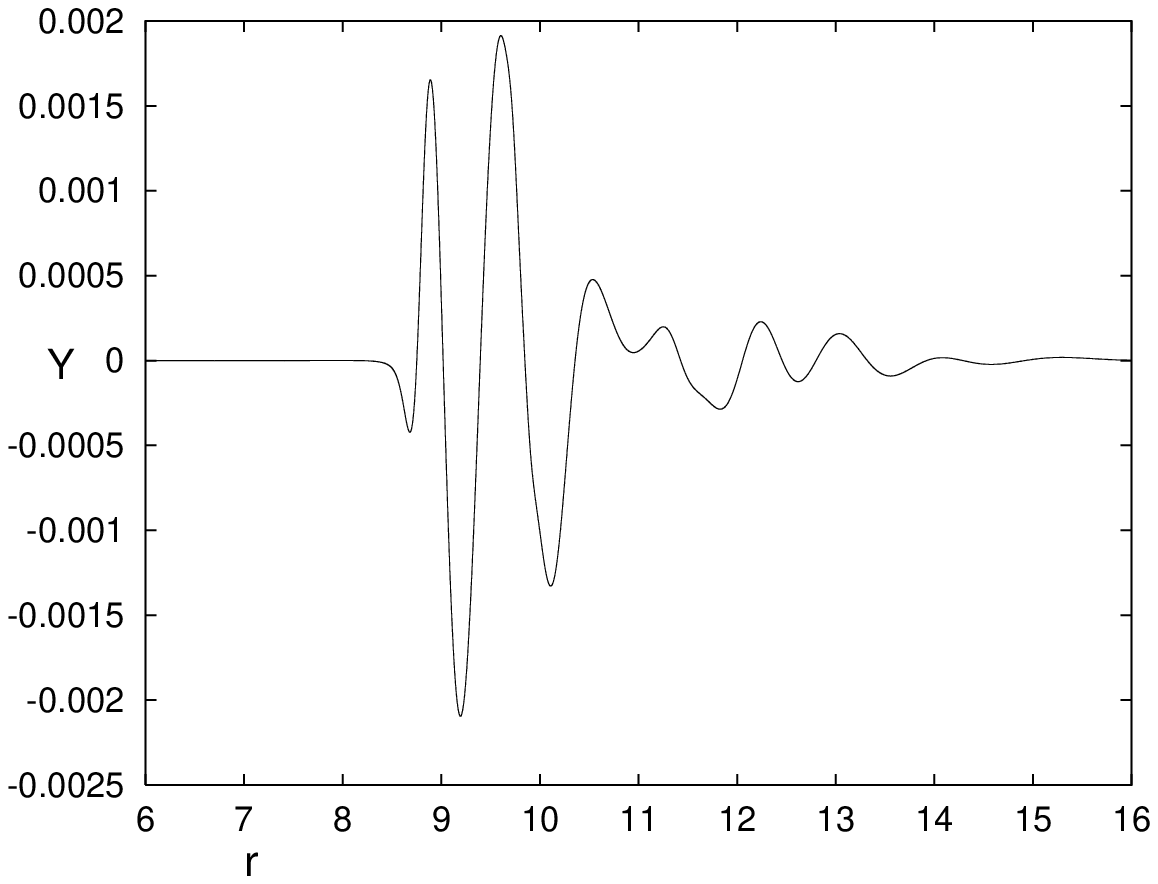}
\caption{\label{fig13} $\phi$ at $t=0.1$. $R=100$ and $N=256000$}
\end{figure}
These numerical results for $R=100$ do not directly address the 
conjecture of \cite{Gary} which is for $R \ge 600$.  However, a 
wall with $R=600$ leads to a wavepacket that is much more narrow
in $\rho$ and therefore requires far more resolution and thus
much more computer memory and time for a simulation that evolves all the
way to black hole formation.  Instead we will evolve such initial data
for a comparatively short time that is nonetheless long enough to
(i) verify that the result is a narrow wavepacket and (ii) show that
the wavepacket has enough energy that its collapse will result in a 
black hole.  Figures 10 and 11 show the results of a run with
$R=600$ and $N=1024000$  evolved to a time of $t=0.1$.  Here $\phi$
is plotted in figure 10 and $Y$ is plotted in figure 11.  Note that the
result is a narrow wavepacket centered near $r=10$, just as one would
expect from the treatment of section III.  Also note that 
$\partial \phi /\partial t \sim 3.8 \times {{10}^3} \phi$.  
For comparison, figures
12 and 13 give respectively $\phi$ and $Y$ for a run with $R=100$
and $t=0.1$.  ($N=256000$ for this run).  Note that the $R=600$
wavepacket has a higher amplitude and shorter wavelength than the
$R=100$ wavepacket.  Thus the $R=600$ wavepacket has more energy within
a shell of smaller thickness than the $R=100$ wavepacket.  Since a black
hole forms from the further evolution of the $R=100$ case, it then 
follows that the evolution of the more energetic wavepacket of the
$R=600$ case will also result in the formation of a black hole. 

\section{Conclusions}

We then see that the initial data of \cite{Gary} when evolved, does not
form a naked singularity, but instead forms a small black hole.  In
hindsight, the reason for this is clear.  The authors of \cite{Gary}
assumed, correctly, that the evolution of the central region could be
described for a while by a perturbation of anti de Sitter space.  However,
they also assumed, incorrectly, that the perturbation is homogeneous.
In the initial data $\phi \propto {r^\beta}$ where $\beta \approx 8.6$.
Thus the initial amplitude for the homogeneous part of the
perturbation is so 
small as to be negligible.  Instead, the perturbation comes from the
wall, {\it i.e.} the transition region between the two vacua.  In the
$\rho$ coordinate, this transition region is very narrow (especially
for walls with large $R$).  Therefore the evolution of 
this initial data is a narrow wavepacket that propagates inward.
In the course of the evolution, the wavepacket becomes sufficiently
concentrated that it is no longer well described by perturbation 
theory.  The numerical simulation then shows that the wavepacket
collapses to form a small black hole.  Thus cosmic censorship is not
violated by the evolution of the initial data given in \cite{Gary}. 

\section{Acknowledgments}

I would like to thank Gary Horowitz for helpful discussions.  This work
was partially supported by a grant from NSERC.

\end{document}